\documentclass[
 reprint,
twocolumn,
superscriptaddress,
longbibliography,nofootinbib,
 amsmath,amssymb,
 aps,
prb,
]{revtex4-2}

\usepackage{CJKutf8}

\usepackage[utf8]{inputenc}
\usepackage{amsmath,amssymb,amsfonts,stmaryrd}
\usepackage{physics}
\usepackage{dsfont} 
\usepackage{graphicx}
\usepackage{xcolor}
\usepackage{graphicx}
\usepackage{cancel}
\usepackage{natbib}
\usepackage{color}
\usepackage{tikz}
\usepackage{mathrsfs}
\usepackage{relsize}
\usepackage{multirow}
\usepackage[linktocpage]{hyperref}
\usepackage[all]{xy}
\hypersetup{colorlinks=true,citecolor=blue,linkcolor=blue, urlcolor=blue, breaklinks=true}

\usepackage[mathscr]{euscript}
\usepackage[scr=boondox]{mathalpha}

\makeatletter
\newcommand\xlabel[2][]{\phantomsection\def\@currentlabelname{#1}\label{#2}}
\makeatother

\newcommand{\Z}{\mathbb{Z}}
\newcommand{\OO}{\text{o}}

\newcommand{\Ox}{\text{o},x}
\newcommand{\Oy}{\text{o},y}

\begin{document}

\title{Electric polarization in Chern insulators:
\\ Unifying many-body and single-particle approaches}

\author{Yuxuan Zhang}
\author{Maissam Barkeshli}

\affiliation{
Department of Physics, Joint Quantum Institute, and University of Maryland,
College Park, Maryland 20742, USA}

\begin{abstract}
Recently, it has been established that Chern insulators possess an intrinsic two-dimensional electric polarization, despite having gapless edge states and non-localizable Wannier orbitals. This polarization, $\vec{P}_{\text{o}}$, can be defined in a many-body setting from various physical quantities, including dislocation charges, boundary charge distributions, and linear momentum. Importantly, there is a dependence on a choice of real-space origin $\text{o}$ within the unit cell. In contrast, Coh and Vanderbilt extended the single-particle Berry phase definition of polarization to Chern insulators by choosing an arbitrary point in momentum space, $\vec{k}_0$. In this paper, we unify these two approaches and show that when the real-space origin $\OO$ and momentum-space point $\vec{k}_0$ are appropriately chosen in relation to each other, the Berry phase and many-body definitions of polarization are equal. 

\end{abstract}

\maketitle

\section{Introduction}

The theory of electric polarization in solids is a classic topic in physics, dating back to early studies of electromagnetism. An important modern advance was the quantum theory of polarization \cite{kingsmith1993,vanderbilt1993electric,resta2007theory}, which determines the electric polarization in terms of the Berry phase of single-particle Bloch states in momentum space. However this approach encounters a problem in the case of Chern insulators, where the Berry connection is not globally well-defined in the Brillouin zone. As is well-known, this is intimately related to the presence of chiral gapless edge states and the impossibility of representing the states in the Chern band in terms of localized real-space Wannier orbitals \cite{vanderbilt2018berry,bernevig2013topological}. Coh and Vanderbilt \cite{coh2009} pointed out that one can define a polarization using the Berry phase definition, however it requires choosing an arbitrary point in momentum space, referred to here as $\vec{k}_{0}$, in order to make sense of the integral of the Berry connection. Altogether, these issues made it unclear to what extent the electric polarization in Chern insulators is a well-defined, physical quantity, and indeed some past literature suggested that electric polarization may not be well-defined for Chern insulators.\footnote{Note that the polarization being discussed is an intrinsically two-dimensional electric polarization. Of course one can always view the 2d system as an effective 1d system and consider the corresponding 1d electric polarization, which gives a popular way to understand the Chern number in terms of a winding number for this 1d polarization.} 

Recent advances \cite{zhang2022pol,zhang2024pol,zhang2024fractionally} have established that one can indeed define an electric polarization, $\vec{P}_{\OO}$, in both integer and fractional Chern insulators. This can be done by focusing on physically well-defined quantities, such as the electric charge in the vicinity of a lattice dislocation, the length-dependence of the total boundary charge, and the linear momentum of the ground state as a function of applied magnetic field. This has the added benefit that the polarization can be defined in an intrinsically many-body fashion in the interacting setting, and does not rely on a single particle formulation. 

These definitions require a choice of origin $\OO$ in the real-space unit cell. This choice of $\OO$ can be easily understood within a classical picture, since the dipole moment within a unit cell must be computed with respect to an origin in real space. Usually, the choice of real-space origin is ignored in discussions of electric polarization in solids. This is because there is often a neutralizing background charge distribution due to the ions in the crystal, in which case the total polarization including the ionic contribution becomes independent of $\OO$. However in many cases of interest, such as in Chern insulators in a magnetic field, charge neutrality is achieved with a metallic gate, which does not have a well-defined polarization (at least in the conventional sense). It therefore is useful to first understand the polarization of the electrons, and then to later include any ionic contribution, which may not fully neutralize the electric charge. 

In the presence of lattice rotational point group symmetries, we can pick $\OO$ to be a high-symmetry point in the unit cell, and the polarization must be quantized to several discrete possible values; as such, it becomes a topological invariant of the crystalline insulator. This topological invariant can be captured from topological quantum field theory methods \cite{manjunath2021cgt,manjunath2020FQH}, which were used extensively in arriving at the results of \cite{zhang2022pol,zhang2024pol,zhang2024fractionally}. When there is only discrete translational symmetry, but no rotational symmetries, the polarization can be captured through topological field theory, but with an unquantized coefficient \cite{song2019electric,manjunath2021cgt}. 

The above results raise the question of understanding the precise relationship between the real-space origin $\OO$ in the electric polarization, and the arbitrary point in momentum space, $\vec{k}_0$, used in the Berry phase definition. It also raises the question of understanding more clearly how the Berry phase definition implicitly depends on a choice of real-space origin. 

In this paper, we address the above questions. We point out that the Berry phase definition always depends implicitly on a choice of position in real space, $\vec{r}_0$. For Chern insulators, there is an additional dependence on a choice $\vec{k}_0$ in momentum space. We then provide a formula (Eq.~\eqref{eq:id1}), which is the central result of this paper, relating $\OO$, $\vec{k}_0$, and $\vec{r}_0$, and which must be satisfied so that the Berry phase definition matches the polarization obtained from the defect charge response, $\vec{P}_{\OO}$. This result can be viewed as providing an unambiguous way to determine $\vec{k}_0$ in terms of bulk response properties and the choice of real-space origin $\OO$. We empirically verify our result using numerical evidence from the arbitrary flux Harper-Hofstadter model on the square lattice and the Haldane model on the honeycomb lattice.

\subsection{Organization of paper}

We organize the rest of the paper as follows. In Sec. \ref{sec:origin} we briefly review some basic facts about the origin-dependence of the electric polarization. In Sec.~\ref{sec:momentumSpace}, we review the Berry phase definition of the polarization $\vec{\mathcal{P}}_{\vec{r}_0,\vec{k}_0}$ calculated in momentum space, and explain how it depends on the parameters $\vec{r}_0$, $\vec{k}_0$ and other important parameters. We also introduce the parameter constraint involving $\vec{r}_0$, $\vec{k}_0$, $L$, and $\OO$ under which $\vec{\mathcal{P}}_{\vec{r}_0,\vec{k}_0}$ is equivalent to $\vec{P}_{\OO}$. In Sec.~\ref{sec:HH} and Sec.~\ref{sec:haldane} we numerically calculate $\vec{\mathcal{P}}_{\vec{r}_0,\vec{k}_0}$  in the square lattice Harper-Hofstadter model and Haldane model and verify that $\vec{\mathcal{P}}_{\vec{r}_0,\vec{k}_0}$ is equivalent to $\vec{P}_{\OO}$ under the parameter constraint.

\section{Origin-dependence of the electric polarization}
\label{sec:origin}

The electric polarization $\vec{P}_{\OO}$, defined modulo $\mathbb{Z}^2$ in units where the electric charge and lattice spacing are set to unity, was defined systematically for Chern insulators in terms of physical response properties of the microscopic system in \cite{zhang2022pol,zhang2024pol}, which also allowed for a non-zero magnetic field.\footnote{Ref. \cite{zhang2022pol,zhang2024pol} defined a quantity $\vec{\mathscr{P}}_{\OO}$, which is related to $\vec{P}_{\OO}$ via $(\mathscr{P}_{\Ox}, \mathscr{P}_{\Oy}) = (P_{\Oy}, -P_{\Ox}) \mod \Z^2$, and for ease of notation referred to $\vec{\mathscr{P}}_{\OO}$ as the polarization.}

Under a change of origin, $\OO \rightarrow \OO + \vec{v}$, Ref. \cite{zhang2022pol} showed how the polarization transforms as:
\begin{align}
\label{Ptrans}
 \vec{P}_{\OO}&\rightarrow \vec{P}_{\OO}-\kappa \vec{v} .
\end{align}
Here $\kappa \equiv \nu - C \frac{\phi}{2\pi}$, where $\nu$ is the charge per unit cell, $C$ is the Chern number, and $\phi$ is the magnetic flux per unit cell. This transformation law can be easily understood in the case where the system is an atomic insulator ($C = 0$). In this case, $\vec{P}_{\OO}$ is the dipole moment per unit cell $\vec{P}_{\OO}=\sum_{j\in \Theta} Q_j \vec{r}_j \mod \mathbb{Z}^2$, where $\Theta$ represents the unit cell, and $\vec{r}_j$ is the position vector of site $j$ relative to $\OO$. Taking $\OO \rightarrow \OO + \vec{v}$ amounts to changing $\vec{r}_j \rightarrow \vec{r}_j - \vec{v}$, giving Eq. \ref{Ptrans}, with $\kappa = \sum_{j\in \Theta} Q_j$ reducing to the charge per unit cell. 

In general, the total polarization of a crystal is decomposed into the electronic and ionic parts,
\begin{align}
\vec{P}_{\OO,\text{tot}}=\vec{P}_{\OO} + \vec{P}_{\OO,\text{ion}}.
\end{align}
Both contributions depend on the choice of origin. Thus the total polarization transforms as
\begin{align}
\vec{P}_{\OO + \vec{v},\text{tot}} = \vec{P}_{\OO, \text{tot}} - (\kappa + \kappa_{\text{ion}}) \vec{v}, 
\end{align}
where $\kappa_{\text{ion}}$ is the ionic charge per unit cell. When the ions completely neutralize the electronic charge, $\nu = -\kappa_{\text{ion}}$. If furthermore $\nu = \kappa$, which occurs if $C$ or $\phi$ vanish, then $\kappa = - \kappa_{\text{ion}}$ and the origin-dependence of $\vec{P}_{\OO, \text{tot}}$ is canceled. It is also possible that the ions completely neutralize the electron charge, $\nu = - \kappa_{\text{ion}}$, but $\nu \neq \kappa$, which occurs if $C \phi/2\pi = - \kappa_{\text{ion}} - \kappa$; in this case, the origin-dependence of $\vec{P}_{\OO,\text{tot}}$ is not canceled. 
In the presence of a metallic gate, it is possible that the ions do not completely neutralize the electron charge in the two-dimensional electron fluid of interest, so $\nu \neq -\kappa_{\text{ion}}$, and the additional neutralizing charge is provided by the gate, which does not have a well-defined polarization. This provides another mechanism for $\kappa \neq -\kappa_{\text{ion}}$ and for $\vec{P}_{\OO,\text{tot}} = \vec{P}_{\OO} + \vec{P}_{\OO,\text{ion}}$ to have a non-trivial dependence on $\OO$. 

We thus see that when the Chern number $C$ and magnetic flux $\phi$ per unit cell are non-zero, we must in general confront the origin-dependence of the polarization. 

\section{Berry phase theory of polarization}\label{sec:momentumSpace}

In this section we review the definition of the polarization in terms of the Berry phase of single-particle Bloch states, and we systematically discuss its dependence on a choice of position in real-space, $\vec{r}_0$ and the choice $\vec{k}_{0}$ in momentum space. 

To be specific, we define the quantity $\vec{\mathcal{P}}_{\vec{r}_0,\vec{k}_0}$, which is the electric polarization as defined by the Berry phase formulation. We will see that for a specific choice of $\vec{r}_0$ and $\vec{k}_\OO$, $\vec{\mathcal{P}}_{\vec{r}_0,\vec{k}_0} = \vec{P}_\OO$. That is, the Berry phase definition agrees with the electric polarization obtained through the physical response properties, such as the defect charges.

\begin{figure}[t]
    \centering
    \includegraphics[width=8cm]{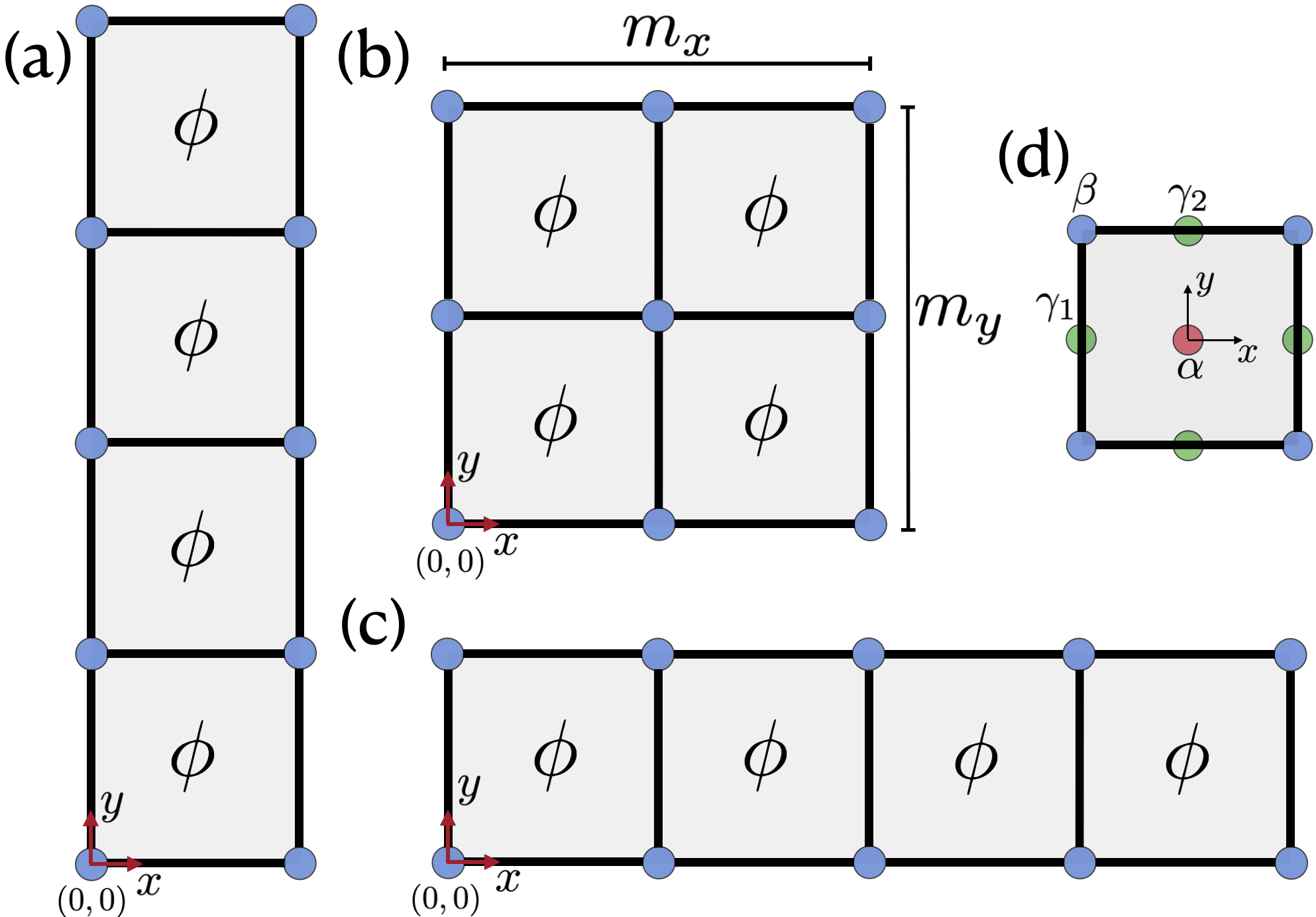}
    \caption{\textbf{(a-c)}Three choices of magnetic unit cell for $\phi=\frac{1}{4}2\pi$. $m_x$ and $m_y$ are the integer linear sizes of the magnetic unit cell in the $x$ and $y$ direction. The $(0,0)$ position is always set to the site at the bottom left corner of the magnetic unit cell. \textbf{(d)} Square lattice unit cell with high symmetry points $\alpha,\beta,\gamma_1,\gamma_2$.}
    \label{fig:magU}
\end{figure}

We let $\vec{a}_x$ and $\vec{a}_y$ be basis vectors of the lattice, which has a flux $\phi = 2\pi p /q$ per unit cell. For simplicity we scale the coordinates so that the lattice spacing is unity in each direction: $|a_x|=|a_y|=1$. Let $\vec{m}_x$ and $\vec{m}_y$ be basis vectors of the lattice obtained by tiling the \textit{magnetic} unit cell. That is, the unit cell generated by $\vec{m}_x$, $\vec{m}_y$ is the magnetic unit cell, which encloses an integer multiple of $2\pi$ flux (see Fig.~\ref{fig:magU}). The reciprocal lattice is defined in the usual manner, and is generated by the reciprocal lattice vectors 
\begin{align}
    \vec{b}_x=-2\pi\frac{Q\vec{m}_y}{S},\qquad
    \vec{b}_y=2\pi\frac{Q\vec{m}_x}{S}.
\end{align}
Here, $Q$ is a matrix representing a $\pi/2$ counterclockwise rotation, and $S = m_x m_y \sin \theta = q \sin \theta$ is the area of the magnetic unit cell, which is $q$ times the area of the elementary unit cell. We denote $\{m_x, m_y, b_x, b_y\}$ as the magnitude of the lattice and reciprocal lattice vectors, and $\theta$ is the angle between $\vec{m}_x$ and $\vec{m}_y$. We assume that $\vec{m}_x$ and $\vec{m}_y$ are chosen such that $m_x$, $m_y$ are integers. 

The lattice momentum $\vec{k}$ can be written in the reciprocal lattice basis as $\vec{k} = k_x \hat{b}_x + k_y \hat{b}_y$. \footnote{Note that $\vec{b}_x$ and $\vec{b}_y$ are not necessarily orthogonal in general.} The real space positions can be expanded as $\vec{r} = r_x \hat{a}_x + r_y \hat{a}_y$, with 
$\hat{b}_x \cdot \hat{a}_x = \hat{b}_y \cdot \hat{a}_y = \sin\theta$ and $\hat{b}_x \cdot \hat{a}_y = \hat{b}_y \cdot \hat{a}_x = 0$.

For a given Hamiltonian, the Berry phase definition of the polarization 
$\vec{\mathcal{P}}_{\vec{r}_0,\vec{k}_0}$, which decompose as $\vec{\mathcal{P}}_{\vec{r}_0,\vec{k}_0}=
\mathcal{P}_{\vec{r}_0,\vec{k}_0,x}\hat{a}_x+\mathcal{P}_{\vec{r}_0,\vec{k}_0,y}\hat{a}_y$.
Below we focus on the $x$-component, $\mathcal{P}_{\vec{r}_0,\vec{k}_0,x}$, as the discussion for the $y$-component is analogous. 
We have the Berry connection
\begin{align}\label{eq:berryphase}
    \mathcal{A}^{(n)}_{\vec{r}_0,x}(\vec{k}) \equiv i \langle u_{n,\vec{k},\vec{r}_0} | \partial_{k_x} | u_{n,\vec{k},\vec{r}_0}\rangle ,
\end{align}
where $|u_{n,\vec{k},\vec{r}_0}\rangle$ are the Bloch states for the $n$th band. As we explain later, the definition of the Bloch states depends on a choice of position in real space, $\vec{r}_0$, which we have made explicit above, and which implies that the Berry connection also depends on this choice. Shifting $\vec{r}_{0}$ amounts to a singular gauge transformation for $\mathcal{A}$. In most discussions in the literature $\vec{r}_0 = (0,0)$ is the canonical choice and it is not explicitly included. We need to track it here because as we will see the Berry phase definition of polarization changes under singular gauge transformations, and therefore changes with $\vec{r}_0$. 

We then define a 1d polarization for each $k_y$:
\begin{align}\label{eq:pblochx}
\mathcal{P}^{\text{Bloch}}_{\vec{r}_0,x}(k_{y}) \equiv  \sum_{n\in\text{occ}} \int_0^{b_x} \frac{dk_x}{2\pi} \mathcal{A}^{(n)}_{\vec{r}_0, x}(\vec{k}),
\end{align}
which includes a sum over occupied bands.
The 2d polarization as defined by the Berry phase formulation is then:
\begin{align}\label{eq:p2d}
    \mathcal{P}_{\vec{r}_0,\vec{k}_0,x} =\int_{k_{0y}}^{k_{0y}+b_y} \frac{\sin \theta d k_{y}}{2\pi} [\mathcal{P}^{\text{Bloch}}_{\vec{r}_0,x}(k_{y}) + \frac{L_x - m_x}{2m_x} n_{\text{fill}}]
\end{align}
The choice $\vec{k}_0=(k_{0x}, k_{0y})$ defines the limits of the integration in the above expression. As pointed out in \cite{coh2009}, when the Chern number vanishes, the polarization is independent of $\vec{k}_0$, however this is not the case for non-zero Chern numbers. The dependence of the above expressions on $\vec{r}_0$ and $\vec{k}_0$ will be discussed in detail in the following section. 
In Eq.~\eqref{eq:p2d}, the term $\frac{L_x - m_x}{2m_x} n_{\text{fill}}$ is a correction derived by Oshikawa and Watanabe \cite{watanabe2018}; we give a brief review of the origin of this term in Appendix ~\ref{sec:derivation}. $n_{\text{fill}}$ represents the number of filled bands, and 
$L_x$ is the number of unit cells (not magnetic unit cells) in the $x-$direction. 

$\mathcal{P}_{\vec{r}_0,\vec{k}_0,x}$ is defined modulo $\frac{1}{m_y}$. Given the same magnetic field $\phi$, there could be many different ways to define the magnetic unit cell, each characterized by different values of $(m_x,m_y)$ (See Fig.~\ref{fig:magU}). It is preferable to pick the one with $m_y=1$ which grants the most information about $\mathcal{P}_{\vec{r}_0,\vec{k}_0,x}$, and similarly pick $m_x=1$ when calculating $\mathcal{P}_{\vec{r}_0,\vec{k}_0,y}$. By choosing the proper magnetic unit cells, we can calculate $\vec{\mathcal{P}}_{\vec{r}_0,\vec{k}_0}\mod \Z^2$.

\subsection{Positional parameters}

In this section, we explain the choice $\vec{r}_0$ in more detail as it appears in the definition of $\vec{\mathcal{P}}_{\vec{r}_0,\vec{k}_0}$. In addition several other real-space quantities also affect $\vec{\mathcal{P}}_{\vec{r}_0,\vec{k}_0}$ in an important way. These include the system size $L_x$, $L_y$ and, in the presence of a magnetic field, a special gauge origin $\bar{\OO}$, which we define below. 

Recall that the Hamiltonian has eigenstates corresponding to the extended wave functions $\psi_{n,\vec{k}}(\vec{r})$, which satisfy $\psi_{n,\vec{k}}(\vec{r} + \vec{m}_i) = e^{i\vec{k} \cdot \vec{m}_i} \psi_{n,\vec{k}}(\vec{r})$. The Bloch wave functions are usually defined as $u_{n,\vec{k}}(\vec{r}) = e^{- i \vec{k} \cdot \vec{r}} \psi_{n,\vec{k}}(\vec{r})$, such that $u_{n,\vec{k}}(\vec{r})=u_{n,\vec{k}}(\vec{r}+\vec{m}_i)$ for $i = x,y$. The real-space position $\vec{r}_0 = (0,0)$ here plays an important role in the definition of the Bloch wave functions. A more general definition is $u_{n,\vec{k},\vec{r}_0}(\vec{r}) = e^{- i \vec{k} \cdot (\vec{r} - \vec{r}_0)} \psi_{n,\vec{k}}(\vec{r})$, which would also be periodic in real space, for any $\vec{r}_0$.

We can also see from the above definitions that $u_{n,\vec{k},\vec{r}_0}(\vec{r})$ is generally not periodic in $k-$space; it satisfies the boundary condition:
\begin{align}\label{eq:bc}
   u_{n,\vec{k}+ \vec{G},\vec{r}_0}(\vec{r}) = e^{-i \vec{G} \cdot (\vec{r}-\vec{r}_0)} u_{n,\vec{k},\vec{r}_0}(\vec{r}), 
\end{align}
where $\vec{G}$ is a reciprocal lattice vector. When $\vec{r}=\vec{r}_0$, $u_{n,\vec{k},\vec{r}_0}(\vec{r}_0)$ is periodic in the Brillouin zone. 

Under a shift, $\vec{r}_0 \rightarrow \vec{r}_0 + \vec{v}_r$, the Bloch state shifts $| u_{n,\vec{k},\vec{r}_0} \rangle \rightarrow e^{i \vec{k} \cdot \vec{v}_r} | u_{n,\vec{k},\vec{r}_0} \rangle$, which amounts to a singular gauge transformation, 
\begin{align}
    \vec{\mathcal{A}}^{(n)}_{\vec{r}_0+\vec{v}}(\vec{k}) \rightarrow  \vec{\mathcal{A}}^{(n)}_{\vec{r}_0}(\vec{k}) - \vec{v}_r ,
\end{align}
which may change $\vec{\mathcal{P}}_{\vec{r}_0, \vec{k}_0}$. 

Next, we observe that the Berry phase definition of the polarization requires that the system be defined on a torus. In this case, there are additional gauge invariant quantities corresponding to the holonomy of the vector potential along non-contractible cycles of the torus. Changing these holonomies amounts to changing $\vec{k}_0$, which therefore changes $\mathcal{P}_{\vec{r}_0, \vec{k}_0}$ for a Chern insulator. We therefore need to pick a reference point for these holonomies and then understand precisely how $\mathcal{P}_{\vec{r}_0, \vec{k}_0}$ changes as this reference point changes. This leads us to the notion of the ``gauge origin" $\bar{\OO}$ defined below. 

Since we are interested in the case of systems with a magnetic field, we consider Hamiltonians $H[A_{ij}^\phi]$ as a function of an applied external vector potential, denoted $A_{ij}^\phi$. We assume that $A_{ij}^\phi$ descends from a continuum vector potential $A^\phi$ in the usual way, where we expand in the basis $A^\phi=A^\phi_x \hat{a}_x+ A^\phi_y\hat{a}_y$. \footnote{Note that given any particular tight-binding Hamiltonian $H$, $A_{ij}^\phi$ and $A_{ij}$ are not a priori determined, since the complex phases of the hoppings may arise from other sources. 
Nevertheless, in a physically realistic system, the tight-binding model is a limit of a system with a well-defined continuum vector potential.}

The continuum vector potential on a torus singles out a particular point in real space (which may not be a site on the lattice), which we refer to as the gauge origin $\bar{\OO}$. $\bar{\OO}$ has the property that the holonomy along the $x$ or $y$ direction, crossing $\bar{\OO}_y$ and $\bar{\OO}_x$ respectively, is trivial. More precisely, we assume the continuum vector potential on a torus has the property that there is a special point $\bar{\OO}$ such that 
\begin{align}
    \oint_{\bar{\OO}_x} A^\phi_y dy = \oint_{\bar{\OO}_y} A^\phi_x dx = 0 ,
\end{align}
where the paths are along the $y$ or $x$ directions of the magnetic lattice and the subscript $\bar{\OO}_i$ indicates position where the two paths crosses on the torus. 

\subsection{Shifting choices}

We now discuss how $\vec{\mathcal{P}}$ changes under a shift of the three parameters $\vec{r}_0$, $\vec{k}_0$, $\bar{\OO}$ and system size $L_x$:
\begin{align}
    \{\vec{k}_0, \bar{\OO}, \vec{r}_0, L_x\}\rightarrow\{\vec{k}_0+\vec{v}_k, \bar{\OO}+\vec{\bar{v}}, \vec{r}_0+\vec{v}_{r}, L_x +\Delta L_x\},
\end{align}
We address these shifts separately.

\subsubsection{Shift of $\vec{r}_0: \vec{r}_0 \rightarrow \vec{r}_0+\vec{v}_{r}$}

As mentioned above, this effectively implements the singular gauge transformation $u_{n,\vec{k}}(\vec{r})\rightarrow e^{i\vec{k}\cdot\vec{v}_{r}}u_{n,\vec{k}}(\vec{r})$, and thus
$\vec{\mathcal{A}}^{(n)}_{\vec{r}_0+\vec{v}_r}(\vec{k}) \rightarrow  \vec{\mathcal{A}}^{(n)}_{\vec{r}_0}(\vec{k}) - \vec{v}_r$. Therefore, 
\begin{align}\label{eq:pblochxshift}
\mathcal{P}^{\text{Bloch}}_{\vec{r}_0,x}(k_{y})
\rightarrow \mathcal{P}^{\text{Bloch}}_{\vec{r}_0,x}(k_{y}) -\frac{n_{\text{fill}}v_{r_x}}{m_x\sin{\theta}} .
\end{align}

Eq.~\eqref{eq:pblochxshift} implies that under $r_{0x}\rightarrow r_{0x}+v_{r_x}$,

\begin{align}
\mathcal{P}_{\vec{r}_0,\vec{k}_0,x}\rightarrow\mathcal{P}_{\vec{r}_0,\vec{k}_0,x}-\frac{n_{\text{fill}}v_{r_x}}{m_x m_y}\mod \frac{1}{m_y}.
\end{align}

\subsubsection{Shift of $\vec{k}_0$: $\vec{k}_0\rightarrow \vec{k}_0+\vec{v}_k$}

In a Chern insulator, the following equality always holds \cite{coh2009}:

\begin{align}
\mathcal{P}^{\text{Bloch}}_{\vec{r}_0,x}(k_{y}+b_y)&=\mathcal{P}^{\text{Bloch}}_{\vec{r}_0,x}(k_{y})- C,
\end{align}
 which implies that under $k_{0y}\rightarrow k_{0y}+v_{k_y}$, 

\begin{align}\label{eq:shiftok}
\mathcal{P}_{\vec{r}_0,\vec{k}_0,x}&\rightarrow \mathcal{P}_{\vec{r}_0,\vec{k}_0,x}-C\frac{v_{k_y}\sin\theta}{2\pi}\mod \frac{1}{m_y} .
\end{align}

\begin{figure}[t]
    \centering
    \includegraphics[width=7.cm]{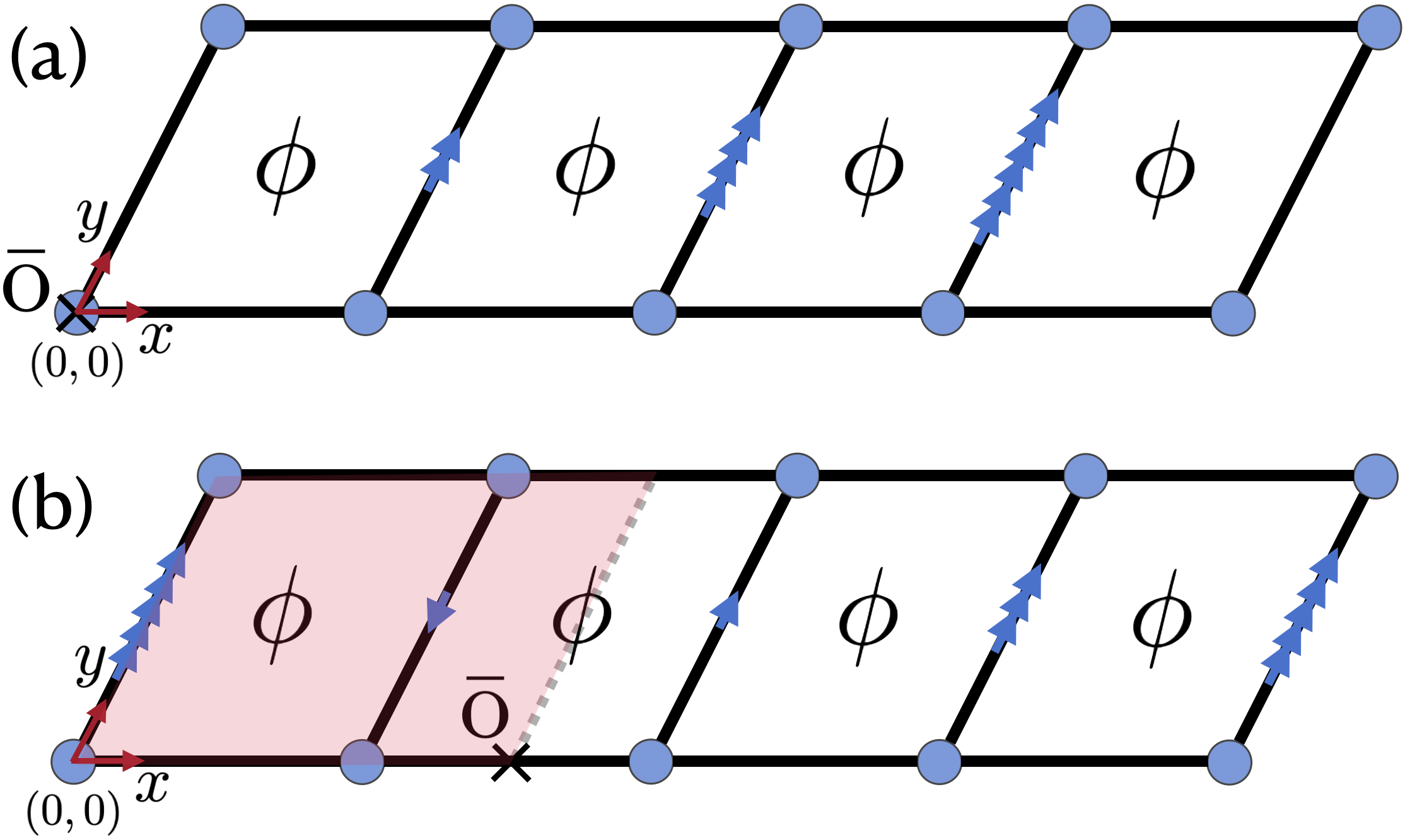}
    \caption{Two gauge choices for $\phi=2\pi/4$ and corresponding magnetic unit cell. Note the $x$ and $y$ axes are not orthogonal. Each blue arrow represents a hopping phase of $e^{i2\pi/8}$. The gauge origins are marked with `$\times$' \textbf{(a)} $\bar{\OO}=(0,0)$, \textbf{(b)} shifting the gauge origin to $\bar{\OO}=(3/2,0)$. The flux in the red shaded region determines the enclosed flux: $\Phi_{\text{enc}}/L_y=3\phi/2$. }
    \label{fig:obar}
\end{figure}

\subsubsection{Shift of gauge origin: $\bar{\OO}\rightarrow\bar{\OO}+\vec{\bar{v}}$}

Changing the gauge origin $\bar{\OO} \rightarrow \bar{\OO} + \vec{\bar{v}}$ amounts to changing the holonomies through the $x$ and $y$ directions of the torus by an amount $\bar{v}_y L_y \phi$ and $\bar{v}_x L_x \phi$, respectively. To see this, note that $\bar{\OO} \rightarrow \bar{\OO} + \vec{\bar{v}}$ implies a change in the vector potential $A^\phi \rightarrow A^\phi + \delta A$ such that
\begin{align}
    \oint_{\bar{\OO}_x + \bar{v}_x} (A_y^\phi + \delta A_y)dy = 0 ,
\end{align}
and similarly for $x \leftrightarrow y$. This implies that 
\begin{align}
\oint_{\bar{\OO}_x + \bar{v}_x} \delta A_y dy = -\Phi_{\text{enc}} ,
\end{align}
where $\Phi_{\text{enc}} = \bar{v}_x L_y \phi$ is the flux in the area enclosed by the two $y$-loops located at $\bar{\OO}_x$ and $\bar{\OO}_x + \bar{v}_x$ (See Fig.~\ref{fig:obar} for an example). Since the magnetic field, set by $\phi$, is left unchanged, we can set $\delta A_y = - \bar{v}_x \phi$.

The above shift has the same effect as shifting the momentum $k_y \rightarrow k_y - \bar{v}_x \phi/\sin \theta$. This is because $A_y^\phi$ enters the Bloch Hamiltonian through the combination $(k_y \sin \theta + A_y^\phi)$. To see this, note that the hopping from site $i$ to $j$ enters the Bloch Hamiltonian through the term 
\begin{align}    [H(k)]_{ij}&=t_{ij}\exp [i(\vec{k}\cdot\vec{r}_{ij}+\int_{\vec{r}_i}^{\vec{r}_i+\vec{r}_{ij}}d\vec{r}\cdot A^\phi)],
\end{align}
where $\vec{r}_{i}$ is the position of the site $i$, $\vec{r}_{ij}$ is the hopping vector from site $i$ to site $j$, and $t_{ij}$ its hopping amplitude. Expanding $\vec{r}_{ij}=r_x\hat{a}_x+r_y\hat{a}_y$,
\begin{align}    [H(k)]_{ij}&=t_{ij}\exp[i((k_xr_x+k_yr_y)\sin\theta+r_xA^\phi_x+r_yA^\phi_y)],
\end{align}
where we used the inner products $\hat{b}_i \cdot \hat{a}_j$ defined above.
Therefore we can conclude that changing the gauge origin $\bar{\OO} \rightarrow \bar{\OO} + \vec{\bar{v}}$ is equivalent to a shift of $\vec{k}_0$ by 
\begin{align}
k_{0y}\rightarrow k_{0y}-\frac{\phi \bar{v}_x}{\sin \theta},
\end{align}
and similarly for $x \leftrightarrow y$. This in turn shifts the polarization as
\begin{align}
\mathcal{P}_{\vec{r}_0,\vec{k}_0,x}&\rightarrow \mathcal{P}_{\vec{r}_0,\vec{k}_0,x}-C\frac{\phi \bar{v}_x}{2\pi}\mod \frac{1}{m_y},
\end{align}
according to Eq.~\eqref{eq:shiftok}. A shift of $\bar{\OO}$ is thus equivalent to a change of $\vec{k}_0$.

\subsubsection{Change of system size: $L_x\rightarrow L_x+\Delta L_x$}

Following Eq.~\eqref{eq:p2d}, this would change $\mathcal{P}_{\vec{r}_0,\vec{k}_0,x}$ as:

\begin{align}
\mathcal{P}_{\vec{r}_0,\vec{k}_0,x}&\rightarrow \mathcal{P}_{\vec{r}_0,\vec{k}_0,x}+\frac{\Delta L_x}{2m_x m_y}n_{\text{fill}}\mod \frac{1}{m_y} .
\end{align}

\subsubsection{Combining all transformations}

Let us express the changes above in terms of $\kappa$: 
\begin{align}
    \kappa \equiv\nu-\frac{C\phi}{2\pi} =\frac{n_{\text{fill}}}{m_x m_y}-\frac{C\phi}{2\pi}.
\end{align}
Note that $\nu$ is the charge per unit cell, $n_{\text{fill}}$ is the charge per magnetic unit cell, and there are $m_x m_y$ unit cells in a magnetic unit cell, so $\nu = \frac{n_{\text{fill}}}{m_x m_y}$. 
To summarize, the total change in $\mathcal{P}_{\vec{r}_0,\vec{k}_0,x}$ under shifts of the parameters discussed above is
\begin{widetext}
\begin{align}\label{eq:shiftO}
    \mathcal{P}_{\vec{r}_0,\vec{k}_0,x}\rightarrow  \mathcal{P}_{\vec{r}_0,\vec{k}_0,x}+\kappa \left( \frac{\Delta{L_x}}{2}-v_{r_x} \right)
    +\frac{C\phi}{2\pi}\left( \frac{\Delta L_x}{2}-v_{r_x}-\frac{v_{k_y}\sin\theta}{\phi}+\bar{v}_x \right) \mod \frac{1}{m_y}.
\end{align}
\end{widetext}

The Berry phase definition of the polarization can only be equal to the many-body definition $\vec{P}_{\OO}$ if they transform the same way under changes of their parameters. In particular, we have 

\begin{align}\label{eq:shiftP}
    P_{\OO + \vec{v},x} = P_{\OO,x}-\kappa v_x \mod 1,
\end{align}
and $\vec{P}_{\OO}$ is independent of all other parameters such as the gauge origin $\bar{\OO}$ and $L_x$. 
Comparing with Eq. ~\eqref{eq:shiftO}, we therefore must have: 
\begin{align}\label{eq:condition1}
    \kappa \left( \frac{\Delta L_x}{2} - v_{r_x} + v_x \right) &= 0 \mod \frac{1}{m_y},
    \nonumber \\
    \frac{C \phi}{2\pi} \left( \frac{\Delta L_x}{2}-v_{r_x}-\frac{v_{k_y}\sin\theta}{\phi}+\bar{v}_x \right) &= 0 \mod \frac{1}{m_y} .
\end{align}
We can take $m_y=1$ to keep the maximal information in $\mathcal{P}_{\vec{r}_0,\vec{k}_0,x}$. Eq.~\eqref{eq:condition1} suggests the following relationship must hold, in order to have $\vec{\mathcal{P}}_{\vec{r}_0, \vec{k}_0} = \vec{P}_{\OO}$:

\begin{figure*}[t]
    \centering
    \includegraphics[width=17cm]{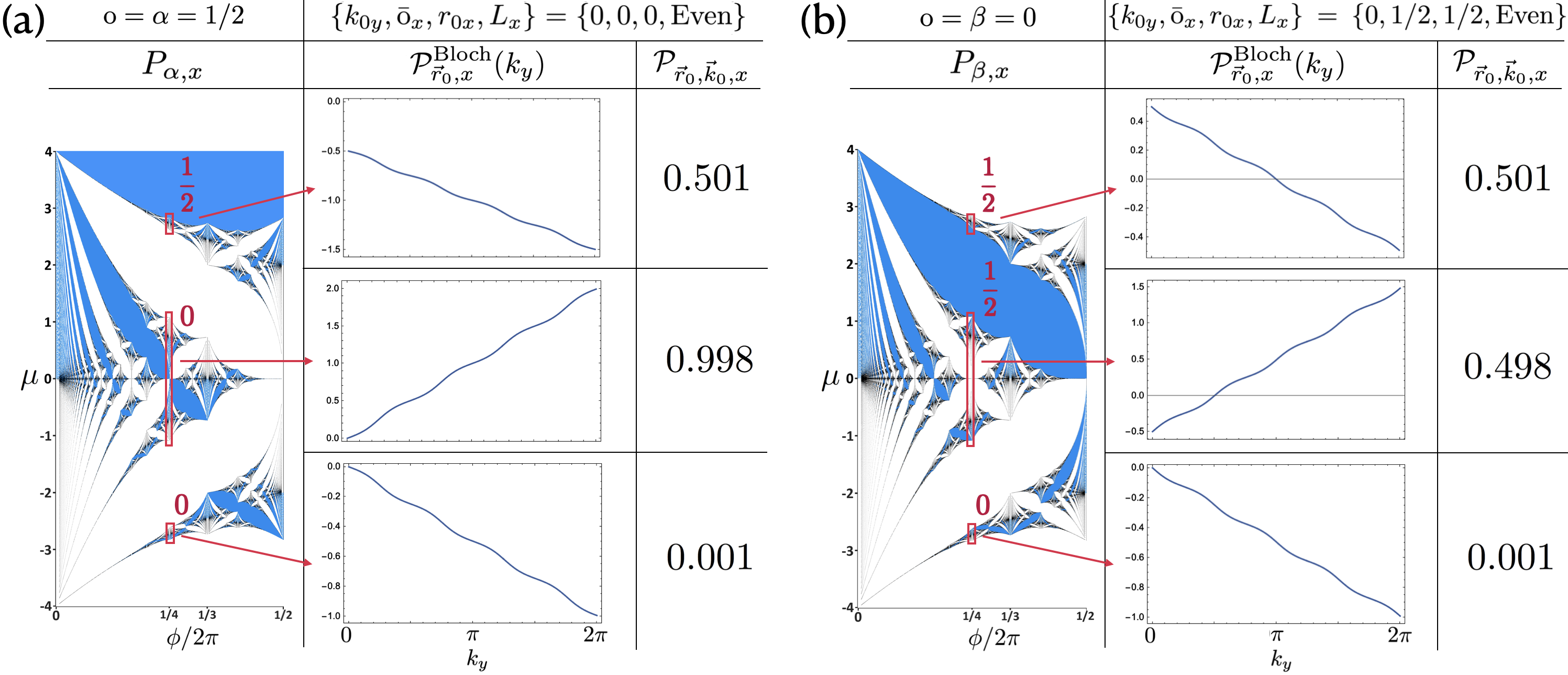}
    \caption{\textbf{(a).} Left panel: Part of the Hofstadter butterfly, colored in with the values of $P_{\alpha,x} = 0,1/2 \mod 1$, originally calculated in 
    \cite{zhang2022pol}. Each white / blue colored space in the butterfly represents the value of ${P}_{\alpha,x}$ where every state with energy below $\mu$ is filled. In each of the red boxes at $\phi = \pi/2$, there is a continuous band of states; fully filling these states gives an insulator with polarization $P_{\alpha,x}$ shown in red. 
    Middle and right panels: $\mathcal{P}^{\text{Bloch}}_{\vec{r}_0,x}$ and $\mathcal{P}_{\vec{r}_0,\vec{k}_0,x}$ are also calculated for these bands with the parameters $\{k_{0y}, \bar{\OO}_x, r_{0x},L_x\} =\{0,0,0,\text{Even}\}$. Note the agreement between $P_{\alpha,x}$ and $\mathcal{P}_{\vec{r}_0, \vec{k}_0, x} \mod 1$.  \textbf{(b).} Same as (a), but comparing $P_{\beta,x}$ with $\mathcal{P}_{\vec{r}_0,\vec{k}_0,x}$, with the parameters $\{k_{0y}, \bar{\OO}_{x}, r_{0x},L_x\}=\{0,1/2,1/2,\text{Even}\}$. }
    \label{fig:HHtot}
\end{figure*}
\begin{widetext}
    \begin{equation}\label{eq:id1}
    \left\{\begin{aligned}
&r_{0x}-\frac{L_x}{2}+c_1=\OO_x \mod 1 &\text{ if } \kappa \neq 0 \\
&\phi\frac{L_x}{2}-\phi r_{0x}+\phi\bar{\OO}_x-k_{0y}\sin\theta+c_2 =0 \mod 2\pi& \text{ if } C \neq 0
\end{aligned}\right.
\end{equation}
\end{widetext}
and similarly for the $y$-components. Here $c_1, c_2$ are constants. 
We test this conjecture numerically in the following sections for various lattice models and confirm that Eq.~\eqref{eq:id1} is indeed correct with $c_1 = 1/2$, $c_2 = 0$. When $\kappa$ or $C$ equals 0, the respective constraint in Eq.~\eqref{eq:id1} is lifted. 

Eq.~\eqref{eq:id1} with $c_1 = 1/2$, $c_2 = 0$ is the central result of this paper. It tells us how to choose the seemingly arbitrary parameters in the Berry phase definition of polarization for Chern insulators in a magnetic field in order to match the result from physical response properties. 
Notably, a similar form to the second condition also appeared in the discussion of \cite{zhang2022pol} when computing $\vec{P}_\OO$ from linear momentum of the ground state and length dependence of the effective 1d polarization. 

\section{Harper-Hofstadter model}\label{sec:HH}

We now calculate $\mathcal{P}_{\vec{r}_0,\vec{k}_0,x}$ numerically in various microscopic models, in order to test the condition Eq. ~\eqref{eq:id1} and determine $c_1$, $c_2$. We first perform this calculation on the HH(Harper-Hofstadter) model \cite{Harper1955,Hofstadter1976}.

For the same $\phi$, there are many ways to define the magnetic unit cell. Fig.~\ref{fig:magU} shows three different choices of magnetic unit cell. Recall that $\mathcal{P}_{\vec{r}_0,\vec{k}_0,x}$ is defined modulo $\frac{1}{m_y}$. For any $\phi$, we can always set up the magnetic unit cell such that $m_y=1$ as in Fig. \ref{fig:magU} C, which retains maximal information in $\mathcal{P}_{\vec{r}_0,\vec{k}_0,x}$. 

Now we calculate $\mathcal{P}_{\vec{r}_0,\vec{k}_0,x}$ for $\phi=\frac{p}{q}=\frac{1}{4}2\pi$, and will later generalize to arbitrary $\phi$. Here, $\vec{k}_0$ is a point defined in $[0,\pi/4)\times[0,2\pi)$. $\vec{r}_0$ and $\bar{\OO}$ are points defined in $[0,4)\times[0,1)$.
The Bloch Hamiltonian is: 

\begin{equation}\hat{H}=
    \begin{pmatrix}
h_0 & e^{ik_x} & 0 &e^{-ik_x}\\
e^{-ik_x} & h_1 & e^{ik_x} & 0\\
0 & e^{-ik_x} & h_2 & e^{ik_x}\\
e^{ik_x}& 0 & e^{-ik_x} &h_3
\end{pmatrix},
\end{equation}
where $h_j=\cos(k_y+(j-\bar{\OO}_x)\phi)$, $j\in\{0,1,2,3\}$ is the $x$ coordinate of the sites. 

\begin{figure}[t]
    \centering
    \includegraphics[width=8.5cm]{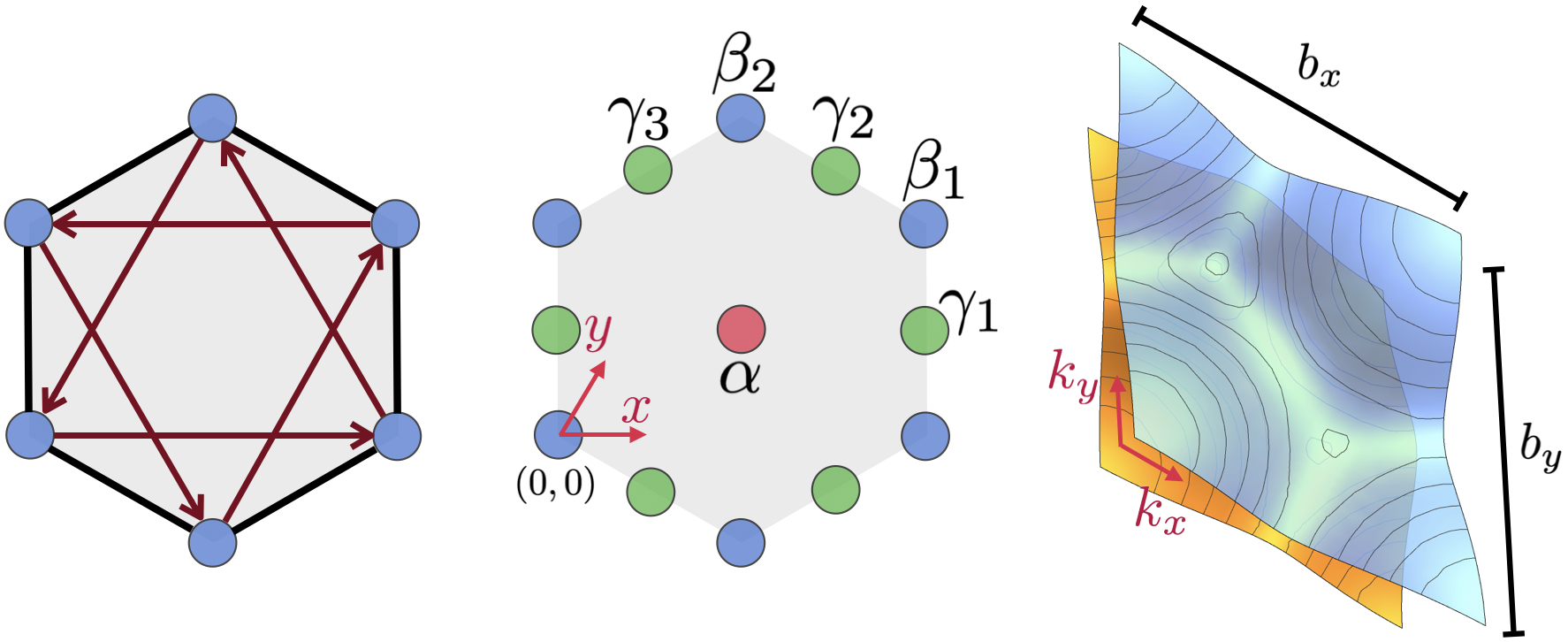}
    \caption{\textbf{Left.} The choice of unit cell in the Haldane model, black lines represent hoppings with coefficient $t_1$ and red lines represent hoppings with coefficient $it_2$. \textbf{Middle}. Maximal Wyckoff positions of the $M=6$ unit cell. The choice of $(0,0)$ position of the unit cell is marked. The high symmetry points $\beta_1$, $\beta_2$ have the same point group symmetry, but are inequivalent under lattice translations; same with $\gamma_i$ points. \textbf{Right.} Band structure of the Haldane model in the first Brillouin zone with parameters $\{t_1,t_2,m\}=\{1,0.1,0.2\}$. The linear size of the reciprocal lattice is $b_x=b_y=\frac{4\pi}{\sqrt{3}}$.}
    \label{fig:haldane_unit}
\end{figure}

\begin{figure*}[t]
    \centering
    \includegraphics[width=16cm]{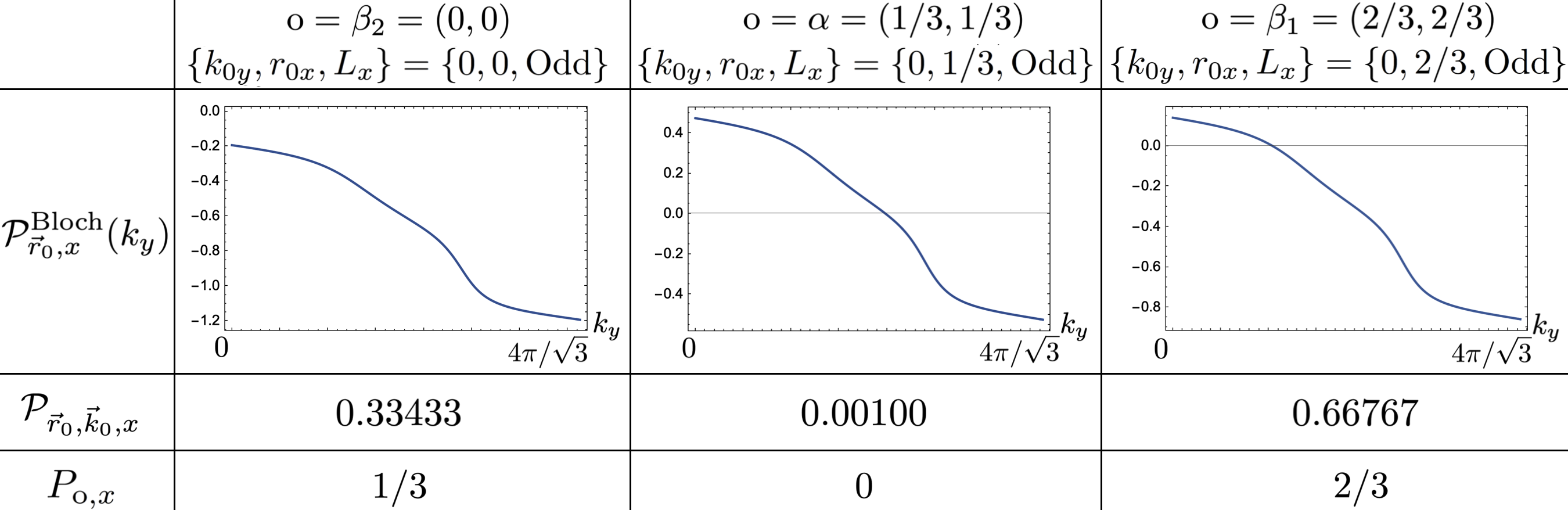}
    \caption{Comparison between $\mathcal{P}_{\vec{r}_0,\vec{k}_0,x}$ and $P_{\OO,x}$ for the Haldane model. }
    \label{fig:haldane_num}
\end{figure*}

We choose the branch of $\mathcal{P}_{\vec{r}_0,\vec{k}_0,x}^{\text{Bloch}}(k_y)$ where the initial value is within the range $-0.5\le\mathcal{P}_{\vec{r}_0,\vec{k}_0,x}^{\text{Bloch}}(k_{0y})<0.5$. A direct numerical calculation of $\mathcal{P}^{\text{Bloch}}_{\vec{r}_0,x}(k_{y})$ and $\mathcal{P}_{\vec{r}_0,\vec{k}_0,x}$ for different sets of positional parameters are shown in Fig.~\ref{fig:HHtot}. The data in the figures suggests that in the $\phi=\frac{1}{4}2\pi$ HH model, under the set of parameters $\{k_{0y}, \bar{\OO}_x, r_{0x},L_x\}=\{0,0,0,\text{Even}\}$, which satisfy the parameter constraint Eq.~\eqref{eq:id1}, we have the equivalence $\mathcal{P}_{\vec{r}_0,\vec{k}_0,x}=P_{\alpha,x}$ up to a $\mathcal{O}(1/L_y)$ correction. Here, $P_{\OO,x}$ is calculated in \cite{zhang2022pol} from dislocation charge and linear momentum, and $\OO=\alpha$ is a high symmetry point that is invariant under $C_{4}$ rotation. $\alpha$ is at the plaquette center (see Fig.~\ref{fig:magU}). For another set of parameters $\{k_{0y}, \bar{\OO}_{x}, r_{0x},L_x\}=\{0,1/2,1/2,\text{Even}\}$ satisfying Eq.~\eqref{eq:id1}, we instead have $\mathcal{P}_{\vec{r}_0,\vec{k}_0,x}=P_{\beta,x}$ up to the same correction. $\OO=\beta$ is another high symmetry point which is at the sites of the square lattice.

In Appendix ~\ref{sec:num}, we demonstrate with more numerical data that under the parameter constraint Eq.~\eqref{eq:id1} we can always establish the equivalence $\mathcal{P}_{\vec{r}_0,\vec{k}_0,x}=P_{\OO,x}$ for the full Hofstadter butterfly, with any $\phi=p/q$ for $q<8$.

\section{Haldane model}\label{sec:haldane}

We now calculate $\mathcal{P}_{\vec{r}_0,\vec{k}_0,x}$ in the Haldane model \cite{haldane1988}, defined on a honeycomb lattice, and establish its connection to $P_{\OO,x}$. The Haldane model has a $C_6$ symmetric unit cell shown in Fig.~\ref{fig:haldane_unit}, where the $x$ and $y$-axis are not orthogonal, and $\phi=0$ so $A^\phi = 0$. The Hamiltonian is
\begin{align}
    \hat{H}=  \begin{pmatrix}
m+h_{nnn}&h_{nn}\\
h_{nn}^*&-m-h_{nnn}
\end{pmatrix}.
\end{align}
where the nearest neighbour hopping $h_{nn}$ and the next nearest neighbour hopping terms $h_{nnn}$ are
\begin{align}
    h_{nn}&=t_1\sum_{i}e^{i\vec{k}'\cdot\vec{w}'_{i}}, \\
    h_{nnn}&=2t_2\sum_{j}\sin(\vec{k}'\cdot\vec{w}'_{j}),
\end{align}
where 
\begin{align}
    &\vec{w}'_i\in\{(0,\frac{1}{\sqrt{3}}),(\frac{1}{2},\frac{-1}{2\sqrt{3}}),(-\frac{1}{2},\frac{-1}{2\sqrt{3}})\}\\
    &\vec{w}'_j\in\{(1,0),(-\frac{1}{2},\frac{\sqrt{3}}{2}),(-\frac{1}{2},\frac{-\sqrt{3}}{2})\}
\end{align}
are the nearest neighbor hopping vector and the next nearest neighbor hopping vector.

Note that $\vec{k}', \vec{w}'$ are defined using the orthogonal coordinates $\hat{x}', \hat{y}'$. However, $\vec{\mathcal{P}}_{\vec{r}_0,\vec{k}_0} $ and $\vec{P}_{\OO}$ are defined in the basis $\hat{x}=\hat{x}'$ and $\hat{y}=\frac{1}{2}\hat{x}'+\sqrt{3}\hat{y}'$, which is not orthogonal. 

The choice of unit cell, high symmetry points, and the first Brillouin zone are shown in Fig.~\ref{fig:haldane_unit}. As derived in \cite{zhang2022pol}, $\vec{P}_{\alpha}=(0,0) \mod \mathbb{Z}^2$ because $\alpha$ has $C_6$ point group symmetry. We can then use Eq.~\eqref{eq:shiftP}
along with the fact that $\kappa\equiv\nu-\frac{C\phi}{2\pi}=1$ to derive the polarization of the $\beta$ points to be
$\vec{P}_{\beta_1}=(2/3,2/3)$ and $\vec{P}_{\beta_2}=(1/3,1/3)$, $\mod \mathbb{Z}^2$.

Now we calculate $\mathcal{P}_{\vec{r}_0,\vec{k}_0,x}$. 
In Fig. \ref{fig:haldane_num}, we numerically test several sets of parameters and show that the equality $\mathcal{P}_{\vec{r}_0,\vec{k}_0,x}=P_{\OO,x}$ is indeed satisfied whenever the parameter constraint Eq.~\eqref{eq:id1} holds.

\section{Discussion}\label{sec:discussion}

In this paper we have discussed two approaches to defining the electric polarization for Chern insulators. One approach \cite{zhang2022pol,zhang2024pol}, based on physical response properties like the lattice dislocation charge, boundary charge, or linear momentum, gives the electric polarization $\vec{P}_\OO$, and depends on a choice of origin $\OO$ in the unit cell in real space. The Berry phase definition gives the quantity $\vec{\mathcal{P}}_{\vec{r}_0, \vec{k}_0}$, which depends on seemingly arbitrary choices in real space and momentum space, $\vec{r}_0$ and $\vec{k}_0$ respectively. In this paper, through a combination of analytical and empirical numerical work, we demonstrated that $\vec{P}_\OO = \vec{\mathcal{P}}_{\vec{r}_0, \vec{k}_0}$, under the condition that $\OO$, $\vec{r}_0$, $\vec{k}_0$, $\bar{\OO}$, and $L_x$ satisfy the parameter constraint Eq.~\eqref{eq:id1}. In particular, this provides a bulk condition for the arbitrary choice $\vec{k}_0$ introduced in \cite{coh2009}. The constants $c_1$ and $c_2$ in Eq.~\eqref{eq:id1} were obtained by fitting the numerical results. It would be interesting to derive values of $c_1$ and $c_2$ analytically in future work, and also to analytically derive the relationship between $\vec{\mathcal{P}}_{\vec{r}_0, \vec{k}_0}$ and the myriad ways of obtaining $\vec{P}_{\OO}$ from dislocation charge, boundary charge, and linear momentum of the ground state, as in \cite{zhang2022pol,zhang2024pol}.

\it Note added \rm-- As this work was being completed, we learned of \cite{gunawardana2025microscopic}, which also addresses the relationship between the polarization obtained from the dislocation charge and the single particle Berry phase definition, although the dependence on the origin $\OO$ and other parameter choices, which is the primary focus of this paper, is not discussed. 

\acknowledgments

We thank Naren Manjunath for discussions, comments on the draft, and collaboration on related work. This work is supported by NSF DMR-2345644 and by NSF QLCI grant OMA-2120757 through the Institute for Robust Quantum Simulation (RQS).

\appendix
\section{Derivation of Eq.~\eqref{eq:p2d}}\label{sec:derivation}

In this section we re-drive the  $\frac{L_x - m_x}{2m_x} n_{\text{fill}}$ contribution in Eq.~\eqref{eq:p2d}, adapted from \cite{watanabe2018}. We consider a 1d ring where $a\equiv m_x$, $L\equiv L_x$ and ignore all origin dependence of $\mathcal{P}$. Generalization to higher dimension is straightforward. 

One common way to define the polarization in 1d is by considering the Hamiltonian $\hat{H}_\theta$ on a ring, as a function of flux (holonomy) $\theta$ through the ring. We assume $\hat{H}_\theta$ has one site per unit cell. The ground state is $|\Phi_\theta \rangle$, and the polarization is usually given as
\begin{align}
\mathcal{P}=\int_0^{2\pi}\frac{d\theta}{2\pi} i\bra{\Phi_{\theta}}\partial_{\theta}\ket{\Phi_{\theta}}\mod 1.
\end{align}
The problem with this definition is that $H_{\theta}$ is in general not periodic in $\theta$, $\hat{H}_{\theta + 2\pi} \neq \hat{H}_{\theta}$, and so $|\Phi_{\theta + 2\pi}\rangle \neq |\Phi_{\theta}\rangle$. In particular, this means we are free to consider any gauge transformation $|\Phi_{\theta}\rangle \rightarrow e^{i\lambda(\theta)} |\Phi_{\theta}\rangle$, where $\lambda$ is not necessarily periodic ($\lambda(\theta+2\pi) \neq \lambda(\theta)$). Under such a transformation, the polarization is not invariant: $\mathcal{P} \rightarrow \mathcal{P} + [\lambda(2\pi) - \lambda(0)]/2\pi$. 

One way to remedy this is to note that for a translationally invariant $\hat{H}_\theta$, the holonomy is spread out uniformly over the ring. For example, $\hat{H}_\theta = \sum_{i} c_i^\dagger c_{i+1} e^{i \theta/L} + H.c.$. Then $e^{i \theta \hat{P}} \hat{H}_{\theta + 2\pi} e^{-i \theta \hat{P}} = \hat{H}_{\theta}$, where 

\begin{align}
    \hat{P}:=\frac{1}{L}\sum_{R/a=0}^{L/a-1} x\hat{n}_x.
\end{align}
where we have decomposed $x=R+r$ where $R = \{0, a, 2a, \dots, L-a\}$ labels the unit cell and $r$ labels the position within the unit cell.
We can pick a gauge where $e^{2\pi i \hat{P}} |\Phi_{2\pi}\rangle = |\Phi_0\rangle$. Then we can make the expression for the polarization gauge invariant by adding an extra term:
\begin{align}\label{eq:gauge_inv_pol}
\mathcal{P}=\int_0^{2\pi}\frac{d\theta}{2\pi} i\bra{\Phi_\theta}\partial_{\theta}\ket{\Phi_\theta} + \text{Im ln}\bra{\Phi_0}e^{2\pi i \hat{P}}\ket{\Phi_{2\pi}}\mod 1,
\end{align}
which is invariant under any gauge transformation $|\Phi_{\theta}\rangle \rightarrow e^{i\lambda(\theta)} |\Phi_{\theta}\rangle$.
Evaluating the above expression in terms of Bloch states will give the additional $(L_x - m_x) n_{\text{fill}}/2m_x$ contribution in Eq.~\eqref{eq:p2d}. 

To evaluate the above expression, it is useful to consider the Hamiltonian:
\begin{align}
\tilde{H}_\theta = e^{i \hat{P} \theta} H_\theta e^{-i \hat{P} \theta},
\end{align}
whose ground state 
\begin{align}\label{eq:Phi_def}
    |\tilde{\Phi}_\theta\rangle = e^{i \theta \hat{P}} |\Phi_\theta\rangle
\end{align}
is periodic, $|\tilde{\Phi}_{\theta + 2\pi} \rangle = |\tilde{\Phi}_{\theta} \rangle $. To carry out this computation we start with the Fourier transform:

\begin{align}
\hat{c}_{k_m^{\theta}} &:= \frac{1}{\sqrt{L/a}} 
\sum_{R/a=0}^{L/a-1} \hat{c}_{x}\, e^{-i k_m^{\theta} x},\\
\hat{\tilde{c}}_{k_m^\theta} &:= \frac{1}{\sqrt{L/a}} 
\sum_{R/a=0}^{L/a-1} \hat{c}_{x}\, e^{-i k_m^{\theta} R} = e^{ik_m^\theta r} \hat{c}_{k_m^\theta},
\end{align}
where we have decomposed $x=R+r$ where $R = \{0, a, 2a, \dots, L-a\}$ labels the unit cell and $r$ labels the position within the unit cell.
Since we have twisted boundary condition, the translation operator satisfy $\hat{\tilde{T}}^L=e^{i\phi\hat{N}}$, and the eigenvalues of $\hat{\tilde{T}}$ are $e^{-ik_m^\theta}$, where $k_m^\theta:=k_m+\theta/L$. $e^{-ik_m}$ is the $m$-th roots of unity.

\begin{figure}[t]
    \centering
    \includegraphics[width=8.5cm]{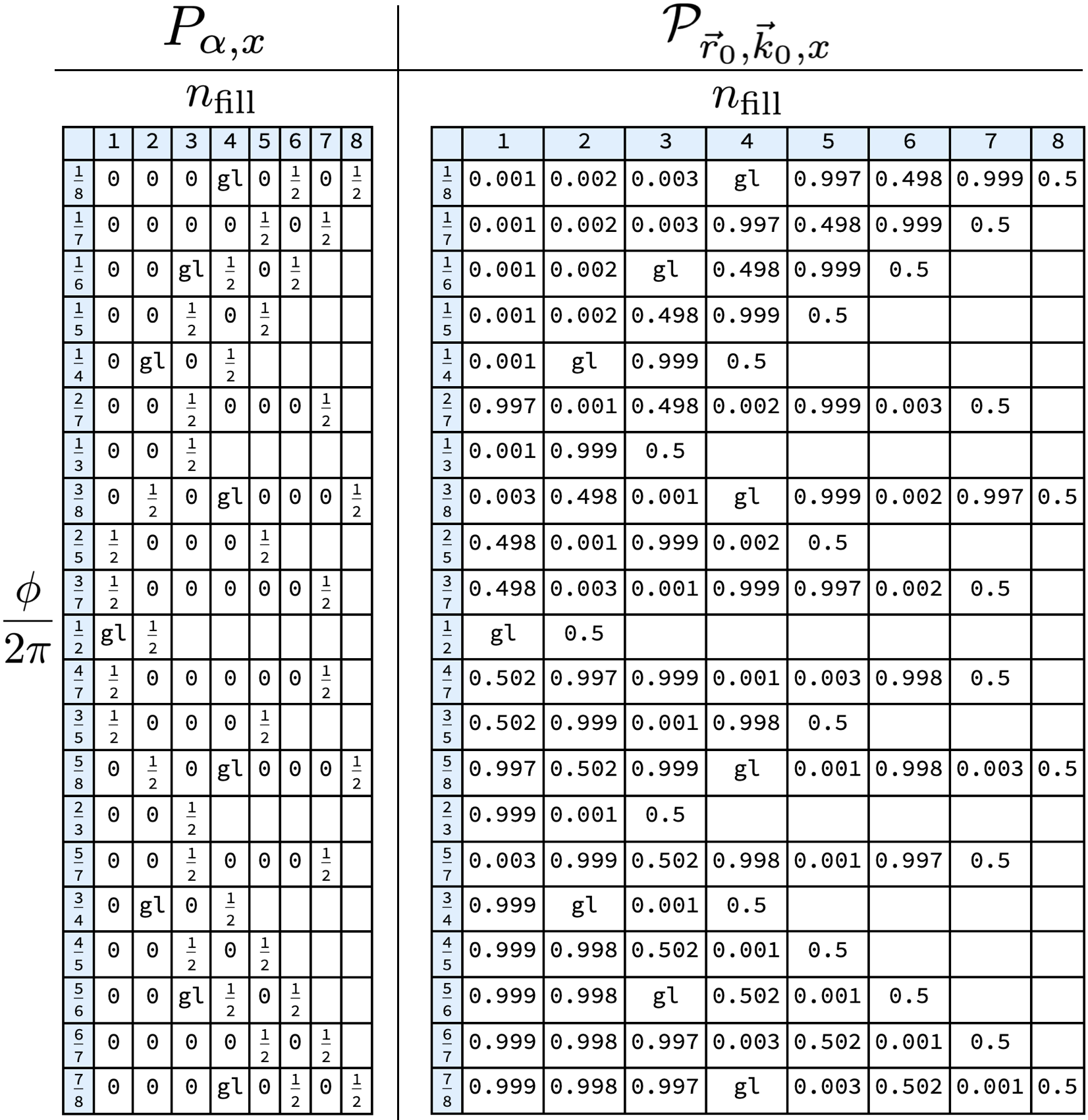}
    \caption{In the square lattice Hofstadter model, the theoretical value of $P_{\alpha,x}$ matches the numerical calculation of $\mathcal{P}_{\vec{r}_0,\vec{k}_0,x}$ up to a $\mathcal{O}(\frac{1}{L_y})$ correction. ``gl" means that filling is gapless. $\mathcal{P}_{\vec{r}_0,\vec{k}_0,x}$ is calculated with $\{k_{0y}, \bar{\OO}_x, r_{0x},L_x\}=\{0,0,0,\text{Even}\}$ which satisfies Eq.~\eqref{eq:id1}. We take $L_x\times L_y=100\times 500$.}
    \label{fig:alphagrid}
\end{figure}

We define the creation operator 

\begin{align} \hat{\tilde{\gamma}}_{n,k_m^\theta}^\dagger
= \int_0^a dr\,\tilde{u}_{n,k_m^\theta}(r)\,\hat{\tilde{c}}_{k_m}^\dagger,
\end{align}
where $n$ is the band index.
The ground state can be expressed as 

\begin{align}\label{eq:theta_state}
    |\tilde{\Phi}_\theta\rangle=e^{-i\frac{L-a}{2a}n_{\text{fill}}\theta}\prod_{m=1}^{L/a}\prod_{n=1}^{n_{\text{fill}}}\hat{\tilde{\gamma}}_{n,k_m^\theta}^\dagger \ket{0}.
\end{align}

When $\theta$ is increased from $0$ to $2\pi$, $k_m^0$ is shifted by $2\pi/L$ to $k_m^{2\pi}=k_{m+1}$. To maintain the $2\pi$ periodicity of $|\tilde{\Phi}_\theta\rangle$, we rearrange the fermion creation operators back to the original ordering, producing a factor $(-1)^{L/a-1}n_{\text{fill}}$ that is canceled with the prefactor of Eq.~\eqref{eq:theta_state}.

With the definition of $|\tilde{\Phi}_\theta\rangle$, we can now also define $|\Phi_\theta\rangle$, lets begin with the Fourier transformations

\begin{align}    \hat{c}_{k_m} \;:=\; \frac{1}{\sqrt{L/a}} 
\sum_{R/a=0}^{L/a-1} \hat{c}_{x}\, e^{-i k_m x},
\end{align}
Since we have periodic boundary condition, $\hat{T}^L=1$, the eigenvalues are $e^{-ik_m}$.

Consider the unitary transformation 

\begin{align}\label{eq:unitary_tx}
    e^{-i\theta\hat{P}}\hat{\tilde{c}}_{k_m^\theta}e^{i\theta\hat{P}}=\frac{1}{\sqrt{L/a}} 
\sum_{R/a=0}^{L/a-1} \hat{c}_{x}\, e^{-i k_m^{\theta} R}e^{i\frac{R+r}{L}\theta}=e^{ik_m^\theta r}\hat{c}_{k_m}.
\end{align}
Using Eq.~\eqref{eq:unitary_tx}, the definition $\tilde{u}_{k}(r)=e^{ikr}u_k(r)$, and the definition of $\ket{\Phi_\theta}$ in Eq.~\eqref{eq:Phi_def}, we can define $\hat{\gamma}_{n,k_m^\theta}$ as:

\begin{align} \hat{\gamma}_{n,k_m^\theta}^\dagger
:= e^{-i\theta\hat{P}}\hat{\tilde{\gamma}}_{n,k_m^\theta}e^{i\theta\hat{P}} = \int_0^a dr\,u_{n,k_m^\theta}(r)\,\hat{c}_{k_m}^\dagger,
\end{align}
The ground state $\ket{\Phi_\theta}$ is then

\begin{align}\label{eq:phi_theta}
\ket{\Phi_\theta}=e^{-i[(L-a)/(2a)]n_{\text{fill}}\theta}\prod_{m=1}^{L/a}\prod_{n=1}^{n_{\text{fill}}}\hat{\gamma}_{n,k_m^\theta}^\dagger \ket{0},
\end{align}
which satisfy Eq.~\eqref{eq:Phi_def}.


Plugging in Eq.~\eqref{eq:phi_theta} to  Eq.~\eqref{eq:gauge_inv_pol}, the polarization is

\begin{widetext}
   \begin{align}
    \mathcal{P} \;&=\;
\frac{L - a}{2a}\,n_{\text{fill}} 
\;+\;
\sum_{n=1}^{L/a}\sum_{\alpha=1}^{n_{\text{fill}}}
\int_{0}^{2\pi}\!\frac{d\theta}{2\pi}
i\bigl\langle 0\bigl|\hat{\gamma}_{n,k_{m}^\theta}\partial_{\theta}\hat{\gamma}_{n,k_{m}^\theta}^{\dagger}\bigr|0\bigr\rangle
=
\frac{L - a}{2a}n_{\text{fill}} 
+
\sum_{n=1}^{L/a}\sum_{n=1}^{n_{\text{fill}}}
\int_{0}^{\frac{2\pi}{a}}\!\frac{dk}{2\pi}
\int_{0}^{a}\!dr
i\,u_{n,k_m^\theta}(r)^{*}\partial_{\theta}u_{n,k_m^\theta}(r)\\
&= \frac{L - a}{2a}n_{\text{fill}}
+
\sum_{\alpha=1}^{n_{\text{fill}}}
\int_{0}^{\frac{2\pi}{a}} \!\frac{dk}{2\pi}
\int_{0}^{a} \!dr
iu_{k}^{\alpha}(r)^{*}\partial_{k}u_{k}^{\alpha}(r)\;\equiv\;
\frac{L - a}{2\,a}\,n_{\text{fill}} \;+\;\mathcal{P}^{\mathrm{Bloch}},
\end{align}
 
\end{widetext}
and we have recovered both contributions in Eq.~\eqref{eq:p2d}.

\begin{figure}[t]
    \centering
    \includegraphics[width=8.5cm]{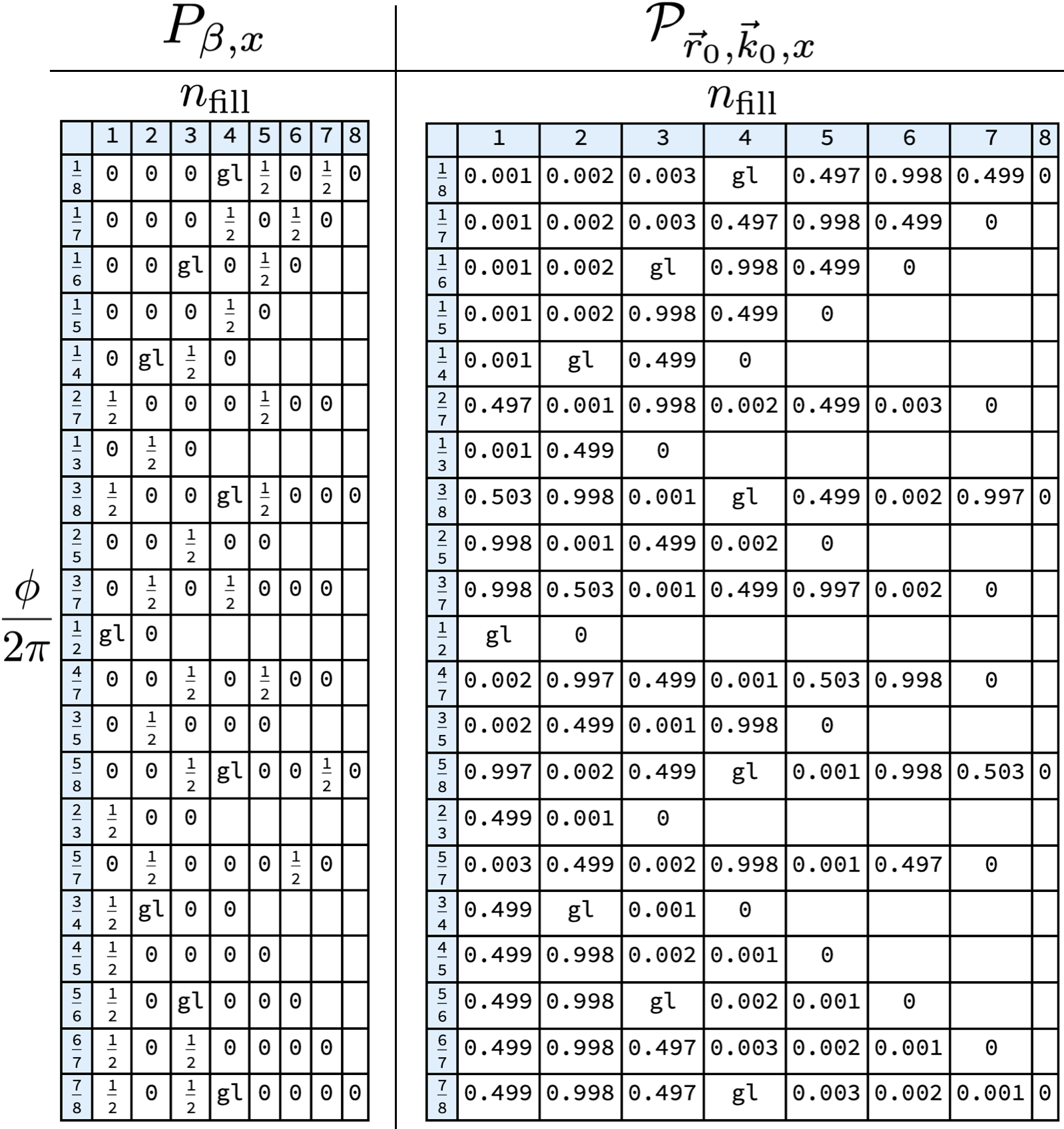}
    \caption{The theoretical value of $P_{\beta,x}$ matches the numerical calculation of $\mathcal{P}_{\vec{r}_0,\vec{k}_0,x}$ up to a $\mathcal{O}(\frac{1}{L_y})$ correction. ``gl" means that filling is gapless. $\mathcal{P}_{\vec{r}_0,\vec{k}_0,x}$ is calculated with $\{k_{0y}, \bar{\OO}_x, r_{0x},L_x\}=\{0,1/2,1/2,\text{Even}\}$ which satisfies Eq.~\eqref{eq:id1}. We take $L_x\times L_y=100\times 500$.}
    \label{fig:betagrid}
\end{figure}

\section{Numerical data: HH model}\label{sec:num}
In this section, we extend the calculation in Sec.~\ref{sec:HH} to any $\phi=\frac{p}{q}$ with $p,q$ coprime and $q\le 8$. We use the finite difference approach to calculate $\mathcal{P}_{\vec{r}_0,\vec{k}_0,x}$. The magnetic unit cell is chosen to have the size $m_x\times m_y=q\times1$ in order to extract $\mathcal{P}_{\vec{r}_0,\vec{k}_0,x}\mod 1$. The magnetic Brillouin zone is discretized into a grid according to

\begin{align}
\nonumber&\vec{k}=(k_x,k_y)\\ 
&k_x=k_{0x}+\frac{2\pi n_x m_x}{L_x}, \quad n_x\in\{0,\dots,L_x-1\}\\ \nonumber
&k_y=k_{0y}+\frac{2\pi n_y m_y}{L_y}, \quad n_y\in\{0,\dots,L_y-1\}
\end{align}

Fig.~\ref{fig:alphagrid} and Fig.~\ref{fig:betagrid} juxtapose the theoretical value of $P_{\OO,x}$ and $\mathcal{P}_{\vec{r}_0,\vec{k}_0,x}$.
When $\phi=p/q$ is promoted to a tunable parameter, the size of the magnetic unit cell may change. In order to extract $\mathcal{P}_{\vec{r}_0,\vec{k}_0,x}\mod 1$, we set $m_y=1$, $m_x=q$. Note that if $q$ is even, $L_x$ is a integer multiple of $m_x$ which is always even. We find that $P_{\OO,x}=\mathcal{P}_{\vec{r}_0,\vec{k}_0,x}$ up to a $\mathcal{O}(\frac{1}{L_y})$ correction whenever $\{k_{0y}, \bar{\OO}_x, r_{0x},L_x,\OO\}$ satisfies the parameter constraint Eq.~\eqref{eq:id1}.

\bibliography{bibliography}

\end{document}